\documentclass[aps,prl,twocolumn,groupedaddress]{revtex4}
\usepackage{graphicx}
\begin{document}

\title{Liquid elasticity length, universal dynamic crossovers and glass transition}

\author{Kostya Trachenko$^{1}$}
\author{V. V. Brazhkin$^{2}$}
\address{$^1$ Department of Earth Sciences, University of Cambridge, UK}
\address{$^2$ Institute for High Pressure Physics, Russia}

\begin{abstract}
We discuss two main universal dynamic crossovers in a liquid that correspond to relaxation times of 1 ps and $10^{-7}$--$10^{-6}$
s. We introduce the concept of liquid elasticity length $d_{\rm el}$. At room temperature, $d_{\rm el}$ is several \AA\ in water
and increases to 0.01 mm in honey and 1 mm in tar. We discuss that on temperature decrease, $d_{\rm el}=d_{\rm m}$ and $d_{\rm
el}=L$ correspond to the two dynamic crossovers, where $d_{\rm m}$ is the medium-range order and $L$ is system size. The second
crossover defines all kinetic aspects of the glass transition whereas ``thermodynamic'' glass transition is realized in the limit
of infinite system size only. One prediction of this picture is the increase of viscosity with the size of macroscopic system,
which we verify by measuring the viscosity of honey.

\end{abstract}

%\pacs{61.43.Fs, 64.70.Pf, 61.20.Lc}

\maketitle

\section{Introduction}

A conceptually simple phenomenon, freezing of liquid into glass, has turned out to be one of the most difficult problems in
condensed matter physics, the problem of glass transition \cite{1,2}. Analyzing the current state of the field, Dyre recently
suggested that glass transition itself is not a big mystery: it universally occurs in any liquid when its relaxation time $\tau$
exceeds the time of experiment at the glass transition temperature $T_g$ \cite{2}. The challenges lie above $T_g$. If we consider
the changes in dynamics in a liquid on lowering the temperature, we find two dynamic crossovers. The first crossover is at high
temperature at liquid relaxation time $\tau\approx 1$ ps, at which dynamics changes from exponential Debye relaxation to
stretched exponential relaxation, SER, $q(t)\propto\exp(-t/\tau)^\beta$, where $q$ is a relaxing quantity and 0$<\beta<$1. This
crossover is universal, i.e. is seen in many systems \cite{3,4,5,6,7,8}. As the temperature is lowered, we find another universal
crossover at $\tau=10^{-7}$--$10^{-6}$ s \cite{9,10}. This crossover also marks the qualitative change in system's dynamics
\cite{9,10,18,19,20,21,22}, and was attributed to the transition from the ``liquid-like'' to the ``solid-like'' behaviour
\cite{10}. Note that although relaxation time at the second crossover is much larger as compared to the first one, it is still
about 9-10 orders magnitude smaller that relaxation time at the glass transition ($\approx 10^3$ s).

One hopes to find useful insights into the problem of glass transition if the origin of the two dynamic crossovers could be
rationalized.

A glass is different from a liquid by virtue of its ability to support shear stress. This suggests that a theory of glass
transition should discuss how this ability changes on lowering the temperature, yet such a discussion is absent from popular
glass transition theories, including free volume and entropy theories \cite{2}. It therefore appears that the most important
physical process in glass transition, stress relaxation mechanism of a liquid, has not been discussed, and is not understood
despite many decades of research.

In this paper, our approach to the problem of glass transition is essentially based on the ability of a liquid to support shear
stress on lowering the temperature. We introduce temperature-dependent liquid elasticity length, which is the range of elastic
interaction in a liquid. Several Angstroms at high temperature, this length grows with relaxation time of the system, increasing
sharply on lowering the temperature. We propose that that the first and second crossovers take place when this length becomes
equal to the values of the medium-range order and system size, respectively. In this picture, we discuss how the second dynamic
crossover is related to the old question of whether glass transition is a thermodynamic or kinetic phenomenon. Finally, we
perform the experimental measurement of viscosity at different values of liquid elasticity length to test the prediction of our
theory that viscosity increases with the size of the macroscopic system.

\section{Liquid elasticity length}

In this section, we introduce the important concept of liquid elasticity length and discuss how it compares with characteristic
lengths in a disordered system. Lets consider relaxation in a liquid under, for example, shear stress. Such a relaxation is the
sequence of elementary localized shear structural rearrangements, local relaxation events (LREs) \cite{2}. A typical shear
relaxation event is shown in Figure 1 (term ``concordant'' in the figure caption is not important here, and will be explained in
the next section). The accompanied structural rearrangement produces elastic shear stress that can propagate through the system.
The important question is how does this stress affect relaxation of other LREs in the system?

\begin{figure}
\begin{center}
{\scalebox{0.7}{\includegraphics{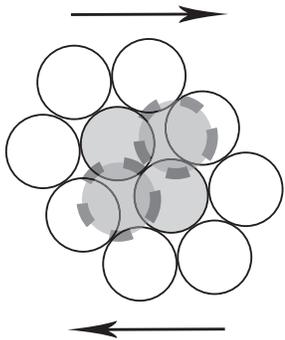}}}
\end{center}
\caption{Orowan's example of a concordant local rearrangement \cite{11}. Solid and dashed lines around the shaded atoms
correspond to initial and final positions of a rearrangement, respectively. Arrows show the direction of external stress.}
\end{figure}

Lets consider how LREs interact elastically, namely how the changes of stresses due to remote LREs affect a given local relaxing
region, shown in the centre in Figure 2. A remote shear LRE, creates the elastic shear waves of different frequencies. Among
these frequencies, the high-frequency wave are present, because the deformation, associated with a LRE, creates a wave with a
length comparable to interatomic separation (see Figure 1), and hence with a frequency on the order of Debye frequency. At high
frequency, larger than the inverse of $\tau$, a liquid can support shear stress just like a solid. Hence the high-frequency shear
waves from all remote LREs propagate the stress and its variations to the central region. However, not all of these contribute to
the increase of stress on the central relaxing region, but only those that are located with the distance $d_{\rm el}=c\tau$ from
the central LRE, where $c$ is the speed of sound. This is because the stresses that arrive to the centre from larger distances
take time longer than $\tau$ to travel, during which the central LRE already relaxes (the time between consecutive LREs is given
by $\tau$). After time $\tau$ a new LRE happens in its place or nearby, and the process repeats.

\begin{figure}
\begin{center}
{\scalebox{0.45}{\includegraphics{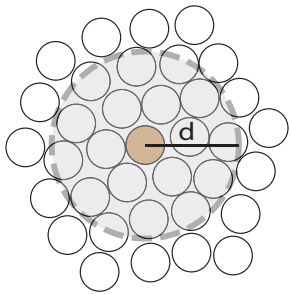}}}
\end{center}
\caption{Illustration of the elastic interaction between local relaxation events. This interaction takes place within the range
$d_{\rm el}$ from the central relaxing regions. Shaded and open circles represent local relaxing regions inside and outside,
respectively, of the interaction sphere.}
\end{figure}

Therefore, $d_{\rm el}$ defines the maximal distance at which two given LREs elastically interact, and can be called {\it liquid
elasticity length}. Because $c$ is on the order of $a/\tau_0$, where $a$ is the interatomic separation of about 1 \AA\ and
$\tau_0$ the oscillation period, or inverse of Debye frequency ($\tau_0=0.1$ ps), we find

\begin{equation}
d_{\rm el}=c\tau=\frac{\tau}{\tau_0}a
\end{equation}

To illustrate the actual values of the introduced elasticity length in real liquids at room temperature, we have calculated
$d_{\rm el}$ from Eq. (1) using the experimental values of viscosity $\nu$ and Maxwell relation $\nu=\tau G_{\infty}$, assuming
$G_{\infty}\approx$ 10 GPa and $c\approx 1000$ m/sec. The results are summarized in Table 1. Starting from 1--1000 \AA\ in
familiar liquids like water or olive oil, $d_{\rm el}$ increases to 0.01 mm in honey and to 1 mm in very viscous tar. In
extremely viscous pitch, $d_{\rm el}=10$ m. An interesting observation here is that for most familiar liquids, $d_{\rm el}$ does
not exceed their typical experimental sizes. The subject of system size will be discussed in this section as well as throughout
this paper.

\begin{table}
\caption{Approximate values of viscosity $\nu$, relaxation time $\tau$ and liquid elasticity length $d_{\rm el}$ at room
temperature.} \vspace{0.5cm}
\begin{tabular}{llll}
\hline \hline
liquid               & $\nu$(Pa$\cdot$s)& $\tau$(sec)            & $d_{\rm el}$\\
\hline
water, olive oil,\\
ethanol, glycerol    & $10^{-3}$--1   &$10^{-13}$--$10^{-11}$  &1--1000 \AA\\
honey                & 10--$10^2$      &$10^{-9}$--$10^{-8}$   &1--10 micron\\
tar                  & $10^4$           &$10^{-6}$             &1 mm          \\
pitch                & $10^8$           &$10^{-2}$             &10 m\\
\hline
\end{tabular}
\end{table}

To discuss how $d_{\rm el}$ compares with characteristic lengths in a liquid, recall that in a perfect crystal, there are two
fundamental lengths, lattice constant $a$ and system size $L$. In a disordered system like liquid, there is an additional length
$d_{\rm m}$, which corresponds to the medium-range order, and is defined by local packing. $d_{\rm m}$ is on the order of 10 \AA,
the characteristic size of decay of structural correlations in a disordered media. $d_{\rm m}$ can weakly depend on the substance
and external parameters (temperature, pressure).

On lowering the temperature, $d_{\rm el}$ crosses all three fundamental lengths in a system. At high temperature, $d_{\rm el}$ is
on the order of interatomic distance $a$ (see Eq. (1)). On lowering the temperature, $\tau$ increases as $\tau=\tau_0\exp(V/kT)$,
and $d_{\rm el}$ quickly increases to $d_{\rm m}$. Because $d_{\rm m}$ is on the order of 10 \AA, we find from Eq. (1) that
$d_{\rm el}=d_{\rm m}$ gives $\tau$ of about 1 ps. This is the first dynamic crossover discussed above. On lowering the
temperature even further, Eq. (1) shows that $d_{\rm el}$ increases to $L$. In liquid relaxation experiments, $L$ is typically
1-10 mm (see discussion below). According to Eq. (1), $d_{\rm el}=L$ gives $\tau=10^{-7}$--$10^{-6}$ s, the second dynamic
crossover. The two crossovers are illustrated in Figure 3.

\begin{figure}
\begin{center}
{\scalebox{0.65}{\includegraphics{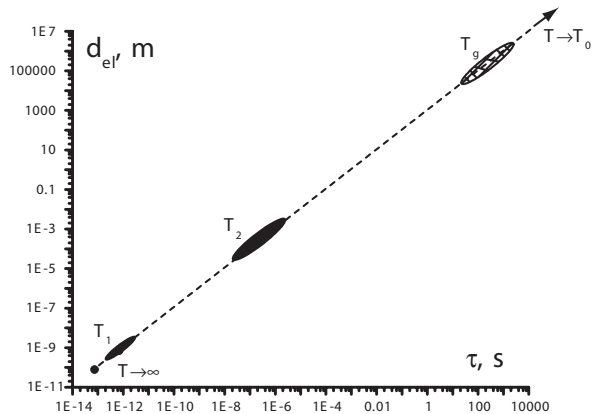}}}
\end{center}
\caption{Elasticity length $d_{\rm el}$ as a function of relaxation time. $T_1$ and $T_2$ correspond to two dynamic crossovers.
$T_g$ and $T_0$ are glass transition, and Vogel-Fulcher-Tamman temperatures, respectively.}
\end{figure}

One of the main proposals in this paper is that the physical reason for the two dynamic crossovers is related to the qualitative
changes of the relaxation mechanism in a liquid on lowering the temperature when $d_{\rm el}$ reaches $d_{\rm m}$ and $L$. A
typical distance between the neighbouring LREs of about 10 \AA, or $d_{\rm m}$. At high temperature, when $d_{\rm el}<d_{\rm m}$,
the elastic interaction between LREs is absent, and LREs take place independently. When $d_{\rm el}=d_{\rm m}$, LREs start
elastically interact, resulting in the first dynamic crossover. When, on lowering the temperature even further, $d_{\rm el}=L$,
all LREs in a system interact with each other. As we discuss below, this gives the second dynamic crossover.

$d_{\rm el}$, introduced in this section, plays the central role in our discussion. In the next section we introduce the elastic
feed-forward interaction mechanism between LREs. The physical importance of $d_{\rm el}$ is that it sets the range of this
interaction.

\section{Elastic feed-forward interaction mechanism}

Some time ago, Orowan introduced ``condordant'' local rearrangement events to discuss creep phenomena \cite{11}. A concordant
local rearrangement is accompanied by a strain agreeing in direction with the applied external stress, and reduces the energy and
local stress (see Figure 1). A discordant rearrangement, on the other hand, increases the energy and local stress. This has led
to a general result that stress relaxation by earlier concordant events leads to the increase of stress on later relaxing regions
in a system. Goldstein applied the same argument to a viscous liquid \cite{12}: consider a system under external stress which is
counterbalanced by stresses supported by local regions. When a local rearrangement to a potential minimum, biased by the external
stress, occurs (a concordant event), this local region supports less stress after the event than before; therefore, other local
regions in the system should support more stress after that event than before \cite{12}.

Lets consider relaxation in a liquid under external shear stress. As argued by Orowan and Goldstein, because an external shear
stress introduces bias towards concordant shear relaxation events, which support less shear stress after relaxation, later LREs
should support more shear stress in order to counterbalance it. If $\Delta p$ is the corresponding increase of stress on a
currently relaxing region and $n$ is the current number of LREs, $\Delta p$ is a monotonically increasing function of $n$.

The increase of stress, $\Delta p$, on a currently relaxing region increases its activation barrier $V$. It has been argued that
$V$ is given by the elastic energy of a surrounding liquid \cite{2,13,14,15}. As discussed by Dyre et al \cite{15}, the energy
needed for an atom to escape its cage at the constant volume is very large because of the strong short-range interatomic
repulsions, hence it is more energetically favourable for the cage to expand, reducing the energy needed for escape. Such an
expansion results in the elastic deformation of the surrounding liquid, hence the activation barrier is given by the work of the
elastic force needed to deform the liquid around a LRE \cite{15}. Because such a deformation does not result in the volume change
of the surrounding liquid (for the displacement field $\bf u$ created by the expanding sphere, div$(\bf u)=0$), it has been
argued that $V$ is given by the shear energy of the surrounding liquid \cite{15}. This result was confirmed by the experimental
measurements of the shear modulus, showing that the activation barrier increases with the shear energy \cite{15}. Because, as
discussed by Orowan and Goldstein, previous LREs reduce stress in the direction ``concordant'' to the external stress (see Figure
1), the increase of shear stress on later rearranging regions consistently increases shear strain on them in the same direction,
increasing shear energy and therefore $V$. The increase of $V$ due to the additional stress $\Delta p$, $\Delta V$, is given by
work $\int \Delta p {\rm d}q$. If $q_a$ is the characteristic volume \cite{15}, $\Delta V=\Delta p q_a$, and we find
$V=V_0+q_a\Delta p$, where $V_0$ is the high-temperature activation barrier.

Because $\Delta p$ is a monotonically increasing function of $n$ and $V=V_0+q_a\Delta p$, we find that $V$ is also a
monotonically increasing function of $n$. This provides the {\it feed-forward interaction mechanism} between LREs, in that
activation barriers increase for later events.

We have recently shown that the two most important signatures of the glass transformation range, stretched-exponential relaxation
and the Vogel-Fulcher-Tamman (VFT) law can be derived on the basis of the feed-forward interaction. For details of calculation
the reader is referred to Refs. \cite{16,17}. In the next two sections we discuss how the feed-forward interaction mechanism
gives rise to the two dynamic crossovers on lowering the temperature when $d_{\rm el}=d_{\rm m}$ and $d_{\rm el}=L$.

The elastic feed-forward interaction mechanism identified here invites the previous discussion about ``cooperativity'' of
molecular motion that becomes operative in a liquid on lowering the temperature. Cooperativity has been discussed intensely in
the area of glass transition \cite{2,yama}. For example, the entropy theory as well as other approaches to the glass transition
assume the existence of ``cooperatively rearranging regions'', ``domains'' or ``clusters'' in a liquid (for review, see Ref.
\cite{yama}), in which atoms move in some concerted way that distinguishes these regions from their surroundings. The physical
origin of cooperativity is not generally known, nor is the nature of the concerted motion. On lowering the temperature, the size
of this region can grow, but does not exceed several nanometers \cite{yama}. A more general view is taken in the Coupling Model
of glass transition, which assumes that on lowering the temperature, the ``primitive'' single-molecule relaxation crosses over to
the cooperative relaxation, in a sense that molecules do not relax as independent, but interact, or ``couple'', in some way
\cite{ngai}.

Here, we do not need to invoke or assume the existence of cooperativity of relaxation as in the previous work \cite{2,yama}. In
our discussion, the elastic interaction between LREs is the necessary feature of relaxation that becomes operative in a liquid on
lowering the temperature: as discussed above, this interaction is absent when $d_{\rm el}<d_{\rm m}$, but becomes operative when
$d_{\rm el}>d_{\rm m}$. Hence in our picture, relaxation is ``cooperative'' in a general sense that LREs are not independent, but
the origin of this cooperativity is the usual elastic interaction. Consequently, instead of the size of a cooperatively
rearranging region discussed previously \cite{2,yama}, we operate in terms of the range over which interactions are elastic in a
liquid. The important quantitative difference between the size of a cooperatively rearranging region and our elasticity length
$d_{\rm el}$ is that the former does not exceed several nanometers \cite{yama}, whereas the latter becomes macroscopic above
$T_g$: if $\tau(T_g)$=10$^3$ sec, $d_{\rm el}$=1000 km at $T_g$, according to Eq. (1).

\section{The first dynamic crossover}

Because, as discussed above, $d_{\rm el}<d_{\rm m}$ at high temperature, the elastic feed-forward interaction is absent, and LREs
do not interact. When LREs are independent, it is easy to derive the expected high-temperature result that relaxation is Debye
(exponential) in time \cite{16}. On the other hand, when $d_{\rm el}$ becomes equal to $d_{\rm m}$ on lowering the temperature,
the elastic feed-forward interaction between LREs becomes operative. We have recently shown that stretched-exponential relaxation
follows as a result \cite{16}. Hence $d_{\rm el}=d_{\rm m}$ marks the crossover from exponential to stretched-exponential
relaxation.

According to Eq. (1), $d_{\rm el}=d_{\rm m}$ gives

\begin{equation}
\tau_1=\frac{d_{\rm m}}{a}\tau_0
\end{equation}

Because $\tau_0=0.1$ ps and $d_{\rm m}/a$ are roughly system- and temperature-independent, Eq. (2) predicts that $\tau_1$ is a
universal parameter. This is consistent with experimental findings \cite{3,4,5,6,7,8}. Because  $d_{\rm m}/a$ is on the order of
10, we find from Eq. (2) that $\tau_1$ is about 1 ps, in good agreement with the experimental value in the 1--2 ps range
\cite{3,4,5,6,7,8}.

\section{The second dynamic crossover}

To discuss the second dynamic crossover, we first discuss how $V$ changes with $d_{\rm el}$. To calculate $V$ as a function of
$d_{\rm el}$, lets consider relaxation induced by an increment of external shear stress. It involves a finite number of LREs, and
we consider, for simplicity, the last LRE to relax to be in the centre of the sphere of radius $d_{\rm el}$ (see Figure 2). The
stress on the central rearranging region increases in order to counter-balance the decreases of stresses due to the previous
remote concordant LREs. These LREs are located within the range of the feed-forward interaction $d_{\rm el}$ (see Figure 2). It
is easy to see that the increase of stress, $\Delta p$, on the central region increases with $d_{\rm el}$: as $d_{\rm el}$
increases, this region needs to counterbalance the reductions of stresses due to an increasing number of remote concordant LREs.
$\Delta p$ can be explicitly calculated as a function of $d_{\rm el}$. Joining the result with $V=V_0+q_a\Delta p$ from section 3
gives \cite{16,17}:

\begin{equation}
V=C_1+C_2\ln(d_{\rm el})
\end{equation}

\noindent where constant $C_1$ and $C_2$ depend on microscopic parameters of the system.

We note here that in Eq. (3), $V$ implicitly depends on temperature through $d_{\rm el}$. Using $\tau=\tau_0\exp(V/kT)$ in Eq.
(1) and combining it with Eq. (3) gives the explicit temperature dependence of $V$. As we have recently shown, this gives the VFT
law for the activation barrier $V=AT/(T-T_0)$ and relaxation time $\tau=\tau_0\exp(A/(T-T_0))$ \cite{17}. Here, the
super-Arrhenius behaviour is related to the increase of the range of the feed-forward interaction, $d_{\rm el}$: as the
temperature is lowered, more LREs are involved in the elastic interaction with a currently relaxing event, increasing its
activation barrier.

We are now ready to discuss the second dynamic crossover. According to Eq. (3), $V$ increases with $d_{\rm el}$ on lowering the
temperature as long as $d_{\rm el}<L$. When $d_{\rm el}=L$, all LREs in the system are involved in the feed-forward interaction.
Hence $d_{\rm el}=L$ marks the transition of the system from being partially to wholly ``elastic'', and should manifest itself as
a qualitative change in the liquid's dynamics. The important aspect of such a change directly follows from Eq. (3): at $d_{\rm
el}>L$, $V\propto\ln(L)$, and is temperature-independent. To be more precise, when $d_{\rm el}>L$, further decrease of
temperature has a weaker effect on $V$, related to, e.g., density increase, but not to the increase of $d_{\rm el}$. As a result,
the system is expected to show a crossover to a more Arrhenius behaviour. This will be discussed in more detail below.

Setting $d_{\rm el}=L$ gives, according to Eq. (1):

\begin{equation}
\tau_2=\frac{L}{a}\tau_0
\end{equation}

Eq. (3) predicts that, similar to $\tau_1$, $\tau_2$ is a universal parameter, independent on temperature or system type. This is
consistent with experimental findings \cite{9,10,19,20,21,22}. Typical values of $L$ used in the experiments are 1--10 mm, which
is dictated mostly by the experimental setup. For example, smaller system sizes can be associated with surface effects, while
larger system sizes can involve temperature gradients and effects of final thermal conductivity. Furthermore, fragile systems of
larger size can not be supercooled without crystallization. Using the range of $L$=1--10 mm and $\tau_0=0.1$ ps, we find from Eq.
(4) that $\tau_2=10^{-7}$--$10^{-6}$ s. This is in good agreement with experimental results which show that for many studied
materials, $\tau_2=10^{-7}$--$10^{-6}$, although exceptions have been noted \cite{22}. Note that at the second $d_{\rm el}=L$
crossover, $1/\tau_2$ has the meaning of the typical values of eigen frequencies of the system.

Experimentally, there is ample evidence for the second dynamic crossover in many systems. Most direct evidence comes from the
sharp kink in the dielectric function \cite{9}. The crossover to the lower slope of relaxation time, with the effect that glass
transition becomes retarded, is observed \cite{18}, in agreement with the prediction from our picture. Other experiments include
NMR relaxation data, which detect a similar dynamic crossover \cite{19}, the crossover in the relaxation of cage sizes in the
positron annihilation experiments \cite{20} and changes of non-ergodicity parameter \cite{21}.

It is interesting to ask what kind of room-temperature liquid has viscosity that corresponds to the second dynamic crossover
$\tau_2=10^{-7}$--$10^{-6}$ s. From Table 1, we observe that $\tau_2$ corresponds to viscosity of $10^3-10^4$ Pa$\cdot$s. This is
much larger than room-temperature viscosity of familiar liquids like water, ethanol, or olive oil, for which $\nu$ is in the
$10^{-3}$--$1$ Pa$\cdot$s range, or honey ($\nu$=10--$10^2$ Pa$\cdot$s). These examples show that although the second crossover
is long above the glass transition, it corresponds to quite high values of viscosity. From Table 1 we find that very viscous tar
($\nu=10^4$ Pa$\cdot$s) has relaxation time close to $\tau_2$ and $d_{\rm el}$=1 mm, comparable with typical experimental system
sizes. This illustrates that the second dynamic crossover $d_{\rm el}=L$ corresponds to a very viscous medium like tar.

\section{Is glass transition thermodynamic or kinetic phenomenon?}

In this section, we discuss how the second dynamic crossover at $d_{\rm el}=L$ is related to the old debate about whether glass
transition should be viewed as thermodynamic or kinetic phenomenon.

As discussed in the previous section, the increase of $d_{\rm el}$ on lowering the temperature gives the VFT law for relaxation
time, $\tau=\tau_0\exp(A/(T-T_0))$ \cite{17}. Because, according to Eq. (1), $d_{\rm el}\propto\tau$, we find that $d_{\rm
el}\propto\exp(A/(T-T_0))$. When $T$ approaches $T_0$, $d_{\rm el}$ diverges, and exceeds any finite size of the system $L$. When
$d\ge L$, all events in the system participate in the feed-forward interaction mechanism, and there is no room for the increase
of $V$ by way of increasing $d_{\rm el}$. As a result, relaxation stops following the VFT dependence, and tends to Arrhenius,
pushing the divergence to zero temperature. We therefore find that only the truly infinite system $L=\infty$ does not have the
second dynamic crossover, and has the divergence of relaxation time at the finite VFT temperature $T_0$. This is illustrated in
Figure 4.

\begin{figure}
\begin{center}
{\scalebox{0.7}{\includegraphics{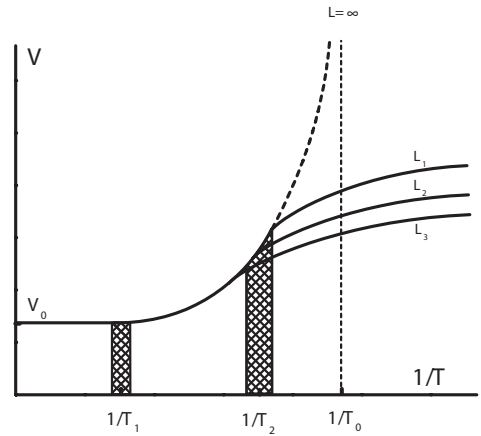}}}
\end{center}
\caption{The effect of the elastic feed-forward interaction on the activation barrier $V$. $V$ increases from its
high-temperature value at the first dynamic crossover at temperature $T_1$ to the second dynamic crossover at $T_2$. After the
second crossover at $d_{\rm el}=L$, $V$ starts saturating to a constant value depending on the system size $L$: the larger $L$,
the larger value of $V$ can be achieved. For finite system size, the divergence of relaxation time takes place at zero
temperature only. For the infinite system, $V$ grows to infinity, resulting in the divergence at the finite VFT temperature
$T_0$.}
\end{figure}

We note that in this picture, no dynamic crossover takes place at $T_g$, consistent with the experimental observations. According
to Eq. (1), $T_g$=10$^3$ s corresponds to $d_{\rm el}$ of 1000 kilometers, which explains why the two dynamic crossovers are seen
long before $T_g$ is reached (see Figure 3). In other words, the absence of a dynamic crossover in the vicinity of $T_g$ is due
to the imbalance between our typical experimental times and sample sizes: at typical experimental time, the elasticity length is
more than 8 orders of magnitude larger than the typical experimental length. In this context, we note that the temperature and
magnitude of the anomalies of thermodynamic properties at $T_g$ are sensitive to cooling rates and observation times. Were the
thermodynamic parameters to be measured at higher cooling rates and shorter experimental times, they would show the anomalies at
temperatures that correspond to relaxation time $\tau_2$. We also note that the thermodynamic anomalies seen at $T_g$ have real
thermodynamic meaning only at $T_0$ which, as discussed above, can be reached only for the infinite system.

It therefore follows from our discussion that liquids on cooling present a unique case when the thermodynamic limit for the phase
transition is not reached for any macroscopic size of the system. Furthermore, system size itself defines the temperature and
existence of the second crossover from liquid-like to solid-like behaviour. In this sense, solidification of a liquid is
dramatically different from behaviour in crystals, where the thermodynamic limit, for practical purposes, is reached for a
sufficiently large (10$^6$) number of atoms, whereas finite size effects come into place at the nanoscale only.

Therefore we conclude that $d_{\rm el}$ plays the role of the order parameter for the glass transition, albeit with unusual
properties: in a finite system, $d_{\rm el}=L$ marks the dynamic crossover without a true thermodynamic transition, whereas
$d_{\rm el}=L$ corresponds to a true thermodynamic transition in the infinite ``thermodynamic'' liquid (see Figure 4).

\section{Experimental results for the system size effect}

The discussed picture of the glass transition makes a specific prediction regarding the effect of system size on viscosity. It
follows from the above discussion that $V$, and hence apparent viscosity, is not a local property, but is governed by the elastic
interaction between local regions within range $d_{\rm el}$. When $d_{\rm el}$ crosses system size at low temperature, viscosity
is defined by interactions between all local regions in the system. According to Eq. (3), $V$ and $\nu$ are independent on system
size for $d_{\rm el}<L$, but increase with $L$ when $d_{\rm el}\ge L$.

It should be noted that the increase of $V$ due to the feed-forward interaction mechanism is not the only possible mechanism of
super-Arrhenius behaviour. For example, in some systems $V$ can noticeably increase due to the density increase on lowering the
temperature. The corresponding contribution to $V$ does not depend on system size, but provides a constant term which can differ
for different types of liquids.

We have chosen honey as an appropriate liquid for our experiment since its elasticity length crosses system sizes in a convenient
temperature range (see below). We have measured viscosity by measuring the falling time of a steel ball using the Stokes equation
with the end and wall correction, $\nu=2gr_b^2(\rho_b-\rho_l)W/9vE$,  where $W=1-2.104(r_b/r_c)+2.09(r_b/r_c)^3-0.95(r_b/r_c)^5$
and $E=1+3.3(r_b/h)$. Here, $v$ is the measured falling velocity, $g$ is the acceleration due to gravity, $\rho_b$ and $\rho_l$
are the densities of the ball (7.7 g/cm3) and the liquid (1.45 g/cm3), $r_b$ and $r_c$ are the radii of the ball ($2r_b=$1.55 mm)
and container, respectively, and $h$ is the container height taken to be the falling distance. We have measured $\nu$ at 306 K,
293 K, 279 K, 254 K and 248 K in two different containers, of $L_1=10$ mm in diameter and $L_1=10$ mm in height, and $L_2=50$ mm
in diameter and $L_2=50$ mm in height. The corresponding values of $\nu_1$ and $\nu_2$ are shown in Table 2. Also shown are the
values of $d_{\rm el}$ calculated from Eq. (1), where $\tau$ is calculated from the Maxwell relation $\nu=\tau G_{\infty}$,
$G_{\infty}\approx 4$ GPa.

\begin{table}
\caption{Viscosities $\nu_1$ and $\nu_2$ measured in containers of length $L_1=10$ mm and $L_2=50$ mm. The calculated values of
$d_{\rm el}$ at five temperatures are also shown.} \vspace{0.5cm}
\begin{tabular}{llll}
\hline \hline
$T$(K) & $d_{\rm el}$ (mm) & $\nu_1$ (Pa$\cdot$s)      & $\nu_2$ (Pa$\cdot$s)\\
\hline
306    & 0.003             & 8.1$\pm$0.5               &8.2$\pm$0.5\\
293    & 0.01              &26$\pm$2                   &25.5$\pm$2\\
279    & 0.08              &210$\pm$15                 &215$\pm$15\\
254    & 45                &(0.9$\pm$0.1)$\cdot$10$^5$ &(1.2$\pm$0.1)$\cdot$10$^5$\\
248    & 110               &(2.1$\pm$0.2)$\cdot$10$^5$ &(2.9$\pm$0.2)$\cdot$10$^5$\\
\hline
\end{tabular}
\end{table}

Consistent with the theoretical prediction, we find that for $d_{\rm el}<L$, $\nu_1=\nu_2$. On the other hand, for $d_{\rm
el}>L$, we observe that $\nu_1<\nu_2$: for $d_{\rm el}$=110 mm, the viscosity is about 40\% larger in the larger container as
compared to the smaller one.

The observed increase of viscosity with system size is an interesting unexpected effect. Since no other theory predicts that
viscosity should increase with system size, we believe that this specific result lends good support to the elastic picture of
glass transition that we discussed.

\section{Summary}

In summary, we have introduced the concept of liquid elasticity length $d_{\rm el}$, and discussed that when $d_{\rm el}$ becomes
equal to the medium-range length and system size on lowering the temperature, two universal dynamic crossovers take place. The
discussion of the second dynamic crossover is directly related to the old debate between the proponents of ``kinetic'' and
``thermodynamic'' views of the glass transition, and we have discussed that glass transition is a kinetic process for any finite
system, and is a thermodynamic transition for the infinite system. Finally, we have experimentally tested the prediction of our
theory that apparent viscosity increases with the size of a macroscopic system.

We are grateful to R. Casalini and M. Roland for discussions and to EPSRC for support.

\end{document}